\documentclass[twocolumn,showpacs,preprintnumbers,amsmath,amssymb,prb,superscriptaddress]{revtex4}
\usepackage{mathrsfs}
\usepackage{graphicx}
\usepackage{dcolumn}
\usepackage{bm}
\usepackage{amsmath}
\usepackage{amsfonts}
\usepackage{txfonts}

\begin{document}

\title{Topological insulators for high performance terahertz to infrared applications}
\author{Xiao Zhang}
\affiliation{Department of Electrical Engineering, Stanford
University, Stanford, CA 94305, USA}
\author{Jing Wang}
\affiliation{Department of Physics, Tsinghua University, Beijing
100084, China}
\author{Shou-Cheng Zhang}
\affiliation{Department of Physics, McCullough Building, Stanford
University, Stanford, CA 94305-4045}

\date{\today}

\begin{abstract}
Topological insulators in the $\mathrm{Bi_2Se_3}$ family have an energy gap
in the bulk and a gapless surface state consisting of a single Dirac
cone. Low frequency optical absorption due to the surface state is
universally determined by the fine structure constant. When the
thickness of these three dimensional topological insulators is
reduced, they become quasi-two dimensional insulators with enhanced
absorbance. The two dimensional insulators can be topologically
trivial or non-trivial depending on the thickness, and we predict
that the optical absorption is larger for topological non-trivial
case compared with the trivial case. Since the three dimensional
topological insulator surface state is intrinsically gapless, we
propose its potential application in wide bandwidth, high
performance photo-detection covering a broad spectrum ranging from
terahertz to infrared. The performance of photodetection can be
dramatically enhanced when the thickness is reduced to several
quintuple layers, with a widely tunable band gap depending on the
thickness.
\end{abstract}


\maketitle

Topological insulators (TIs) are a new state of quantum
matter with an insulating bulk gap and gapless edge or surface
states interesting for condensed matter physics, material science
and electrical
engineering\cite{Qi2009a,bernevig2006d,koenig2007,fu2007a,hsieh2008,Zhang2009,Xia2009,chen2009}.
The two-dimensional (2D) TI, with quantum spin Hall (QSH) effect has
been predicted and observed in $\mathrm{HgTe/CdTe}$ quantum
well\cite{bernevig2006d,koenig2007}. Recently, 3D TI such as
$\mathrm{Bi_2Se_3}$ and $\mathrm{Bi_2Te_3}$ were theoretically
predicted to have bulk energy gap as large as $0.3$eV, and gapless
surface states consisting of a single Dirac
cone\cite{Zhang2009,Xia2009}. The angle-resolved photoemission
spectroscopy (ARPES) observed such linear Dirac spectrum dispersing
from the $\Gamma$ point in both of these
materials\cite{Xia2009,chen2009}. $\mathrm{Bi_2Se_3}$ and
$\mathrm{Bi_2Te_3}$ are stoichiometric rhombohedral crystals with
layered structure consisting of stacked quintuple layers (QLs), with
relatively weak Van der waals coupling between QLs (each QL is about
1nm thick). Therefore high quality thin films have been successfully
grown on silicon and silicon carbide substrates via molecular beam
epitaxy\cite{zhangyi2010,liyaoyi2010,li2010}, layer by layer, which
enables further scientific study and applications integratable with
today's electronics. These materials have also been grown by
Au-catalyzed vapor-liquid-solid\cite{kong2010a} and catalyst-free
vapor-solid chemical vapor deposition methods\cite{kong2010b} on
silicon, silicon dioxide and silicon nitride substrates. The surface
states of such thin film have been predicted
\cite{Zhang2009,Linder2009,Liu2010} and observed to open a gap when
they are thinner than 6 QLs\cite{zhangyi2010}. Up to now, few study
on optical properties of the topological surface states has been
reported. On the other hand, the single Dirac cone on the
$\mathrm{Bi_2Se_3}$ surface can be imagined as $1/4$ of
graphene\cite{novoselov2005}, it is straightforward to study the
optical properties and relevant applications of TI in analogy to the
optoelectronics applications of
graphene\cite{Nair2008,xiafn2009,Neto2009}.

Starting from an effective $\mathbf{k}\cdot\mathbf{p}$ Hamiltonian and ignoring intra-unit cell dipole contribution to absorbance,
we show that the low energy optical absorbance of thin film
$\mathrm{Bi_2Se_3}$ thicker than 6 QLs is a universal quantity
$\pi\alpha/2$, which does not depend on the photon energy or
chemical composition of the material, where $\alpha$ is the fine
structure constant. This originates from Dirac nature of the 2D
topological surface states. Furthermore, the optical transitions
from the valence to conduction surface bands depend solely on the
spin-states in contrast to the conventional semiconductors. When the
thickness of such thin film is less than 6 QLs, a gap is opened up
for the surface states around the $\Gamma$
point\cite{Linder2009,Liu2010,Lu2010,zhangyi2010}, and the resulting
2D insulator can either be topologically trivial, or non-trivial,
depending the thickness of the film\cite{Liu2010}. We show here that
the optical absorbance near the band edge is either smaller or
larger than $\pi\alpha$, depending on it is conventional insulator
or 2D QSH insulator. This suggests an optical mean to measure its
topological nature other than transport\cite{koenig2007}. With such
strong absorbance, the thin film $\mathrm{Bi_2Se_3}$ may be a
promising candidate for high performance photodetector in the
terahertz (THz) to infrared frequency range. The gapless 3D TI can
cover this full spectrum just by a single device with high
signal-to-noise ratio (SNR) comparable to multiple structures using
the conventional photodetecting material - bulk
$\mathrm{Hg_{1-x}Cd_{x}Te}$\cite{dornhaus1976,
hamamatsu2004,rogalski2003} with different fraction $x$. The SNR can
be further enhanced to be about 15 times better with reduced
thickness of $\mathrm{Bi_2Se_3}$ less than 6 QLs. Graphene has been
recently demonstrated as a THz photodetector\cite{xiafn2009}. With
almost the same high absorbance\cite{Nair2008}, TI photodetector
proposed here has many advantages. The prominent one is a tunable
surface band gap which is easily achieved by reducing the thickness,
which is necessary in many optoelectronics
applications\cite{Ryzhii2008}.

\section{Effective model for thin film TI} 3D TI like
$\mathrm{Bi_2Se_3}$ is characterized by a surface state consisting
of a single Dirac cone, where the spin points perpendicular to the
momentum\cite{Zhang2009}
\begin{eqnarray}
\mathcal{H}_{surf} =
A_2\left(k_y\sigma^x-k_x\sigma^y\right),
\end{eqnarray}
with $A_2=\hbar v_F$, $v_F$  is the Fermi velocity and
$\sigma^{x,y}$ are Pauli matrices describing the spin. When the
thickness of a film is reduced, overlapping between the surface
state wave functions from the upper and lower surfaces of the film
becomes non-negligible, and hybridization between them will induce a
gap at Dirac point to avoid crossing of bands\cite{Lu2010}.
Furthermore, the 2D energy gap is predicted to oscillate between the
ordinary insulating gap and QSH gap as a function of
thickness\cite{Liu2010}. The effective model of the quasi-2D system
can be derived from the $\mathbf{k}\cdot\mathbf{p}$ Hamiltonian for 3D TI\cite{Liu2010b},
\begin{equation}\label{3D}
\mathcal{H}^{3D}_0 = \left(
\begin{array}{cccc}
\mathcal{M}(\mathbf{k}) & A_1k_z & 0 & A_2k_-\\
A_1k_z & -\mathcal{M}(\mathbf{k}) & A_2k_- & 0\\
0 & A_2k_+  & \mathcal{M}(\mathbf{k}) & -A_1k_z\\
A_2k_+ & 0 & -A_1k_z & -\mathcal{M}(\mathbf{k})
\end{array}
\right)+\epsilon_0(\mathbf{k}),
\end{equation}
with $k_{\pm}=k_x\pm ik_y$, $\epsilon_0(\mathbf{k})=C_0+D_2
(k_x^2+k_y^2)+D_1k_z^2$,
$\mathcal{M}(\mathbf{k})=M_0+B_2(k_x^2+k_y^2)+B_1k_z^2$, and the
four basis of the 3D Hamiltonian are denoted as
$|P1^+_z,\uparrow\rangle$, $|P2^-_z,\uparrow\rangle$,
$|P1^+_z,\downarrow\rangle$, $|P2^-_z,\downarrow\rangle$ with the
superscript $\pm$ standing for even and odd parity and
$\uparrow(\downarrow)$ for spin up (down). When the TI film is thin,
$k_z$ is no longer a good quantum number, by replacing
$k_z\rightarrow-i\partial_z$ and with the in-plane momentum
$\mathbf{k}_{\parallel}$ good quantum numbers, then 3D model can be
expressed as
$\mathcal{H}^{3D}_0(\mathbf{k}_{\parallel},-i\partial_z)\equiv\mathcal{H}^{3D}_0(\mathbf{k}_{\parallel}=0,-i\partial_z)+\delta\mathcal{H}^{3D}_0$,
where $\mathcal{H}^{3D}_0(\mathbf{k}_{\parallel}=0,-i\partial_z)$ is
block diagonal and can be solved exactly, then 2D thin film model
can be obtained by projecting $\mathcal{H}^{3D}_0$ on the
eigenstates of
$\mathcal{H}^{3D}_0(\mathbf{k}_{\parallel}=0,-i\partial_z)$ (denoted
as $|1,\uparrow\rangle$, $|2,\uparrow\rangle$, $|1,
\downarrow\rangle$, $|2, \downarrow\rangle$),
\begin{equation}\label{BHZ}
\mathcal{H}^{2D}_0 = \left(
\begin{array}{cccc}
\tilde{\mathcal{M}}(\mathbf{k}_{\parallel}) & 0 & 0 & \tilde{A}_2k_-\\
0 & -\tilde{\mathcal{M}}(\mathbf{k}_{\parallel}) & \tilde{A}_2k_- & 0\\
0 & \tilde{A}_2k_+  & \tilde{\mathcal{M}}(\mathbf{k}_{\parallel}) & 0\\
\tilde{A}_2k_+ & 0 & 0 &
-\tilde{\mathcal{M}}(\mathbf{k}_{\parallel})
\end{array}
\right)+\tilde{\epsilon}_0(\mathbf{k}_{\parallel}),
\end{equation}
it is nothing but the model of Bernevig, Hughes and Zhang (BHZ) for
the 2D QSH insulator{\cite{bernevig2006d,Liu2010,Lu2010}} with
$\tilde{\epsilon}_0(\mathbf{k}_{\parallel})=\tilde{C}_0+\tilde{D}_2
(k_x^2+k_y^2)$,
$\tilde{\mathcal{M}}(\mathbf{k}_{\parallel})=\tilde{M}_0+\tilde{B}_2(k_x^2+k_y^2)$,
and $\tilde{C}_0$, $\tilde{D}_2$, $\tilde{M}_0$, $\tilde{B}_2$ and
$\tilde{A}_2$ are parameters which are thickness dependent. The
energy dispersion is
$\epsilon_{\pm}(\mathbf{k})=\tilde{\epsilon}_0(\mathbf{k})\pm
\sqrt{\tilde{A}_2^2
k^2_{\parallel}+\tilde{\mathcal{M}}(\mathbf{k}_{\parallel})^2}$,
which is double degenerate and gives a gap of $2|\tilde{M}_0|$ at
$\Gamma$ point. As has been experimentally demonstrated in Ref.~9, a finite surface gap is induced when the TI film
of $\mathrm{Bi_2Se_3}$ is less than 6 QLs.
\begin{figure}
\begin{center}
\includegraphics[width=2.5in,clip=true]{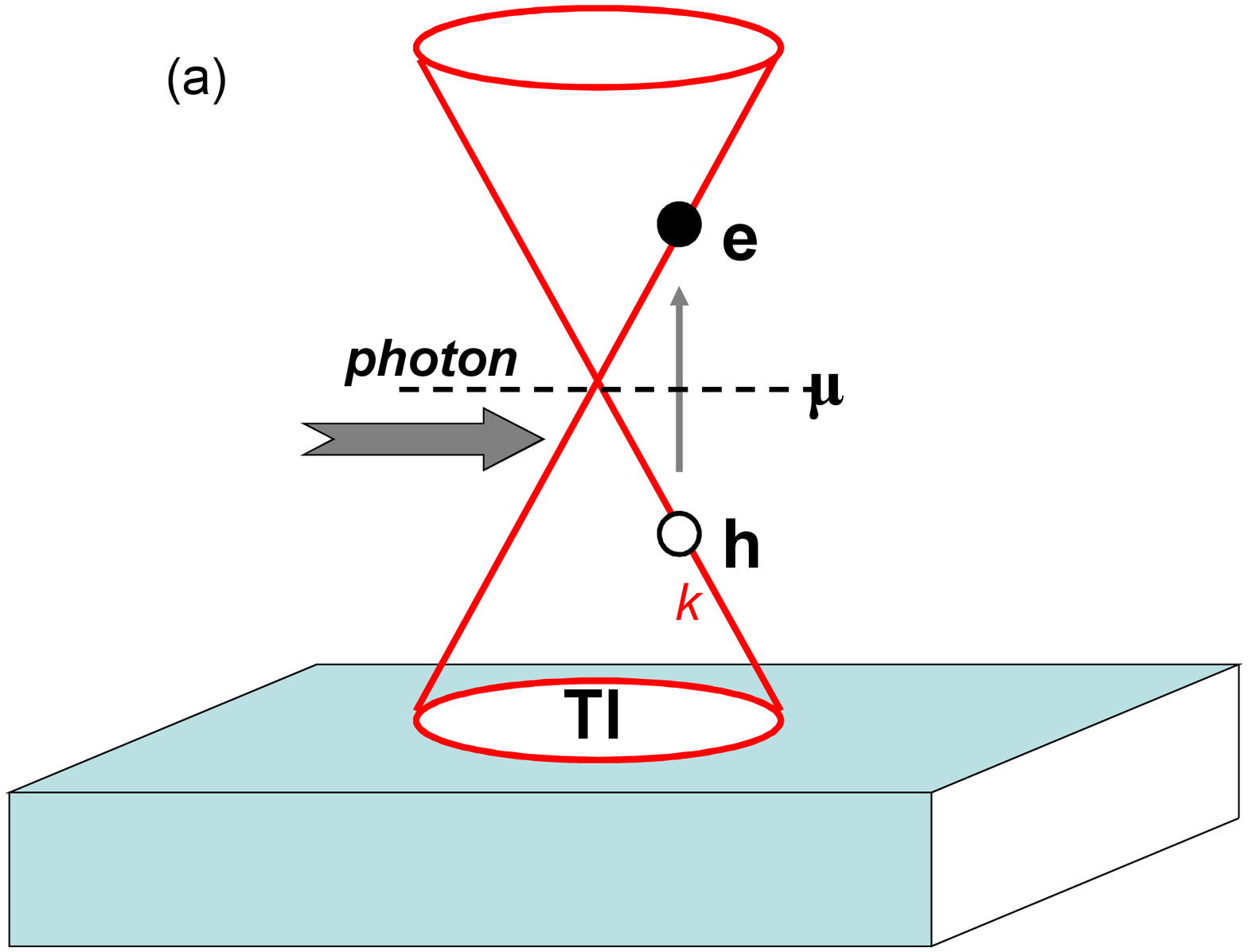}
\includegraphics[width=2.5in,clip=true]{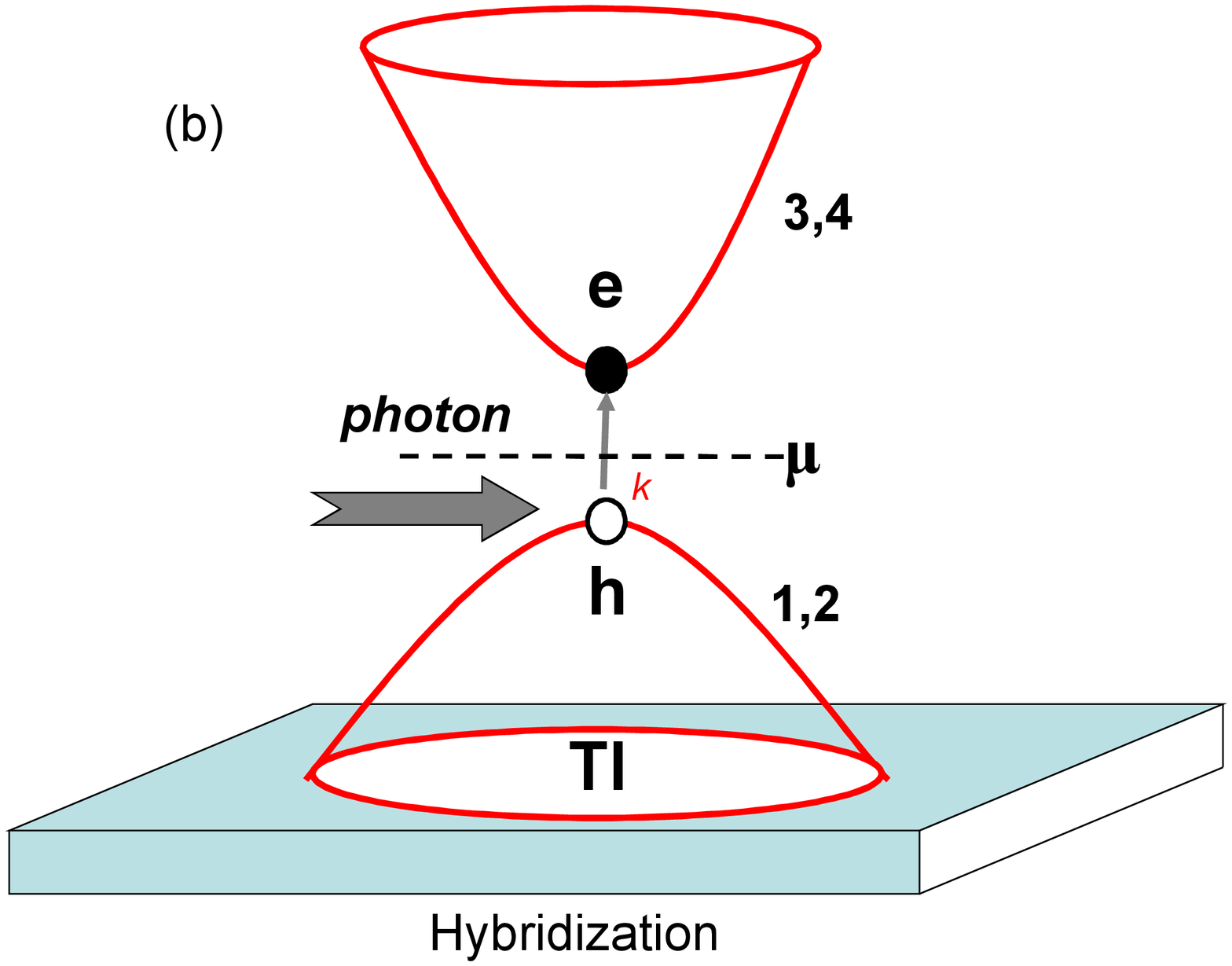}
\caption{\textbf{Optical absorption of surface states.} The particle
and hole pair generation during photon absorption for (a) 3D
and (b) 2D TI surfaces.}
\end{center} \label{fig1}
\end{figure}

\section{Optical absorbance and selection rules}
Coupling to external electromagnetic field $\bf A$ is described by the model Hamiltonian (\ref{BHZ}), with the
replacement of $\bf k$ by ${\bf k}-\frac{e}{c} {\bf A}$. As in the
case of graphene, the massless  Dirac spectrum leads to the
universal optical absorbance
$\mathcal{P}_{\textrm{graphene}}=\pi\alpha$\cite{Nair2008}. Such
optical properties being defined by the fundamental constants
originates from the two-dimensional nature and gapless electronic
spectrum of graphene. This suggests the universal absorbance should
also be true for the helical surface states of TI. When the
thickness of $\mathrm{Bi_2Se_3}$ thin film exceeds 6 QLs, the low
energy optical absorption of its gapless surfaces can always occur
with photon energy ranging from $0$ to $0.3$eV, where the high energy
cut-off is the bulk band gap. Within the dipole moment
approximation, we can write down the interacting Hamiltonian due to
the optical transitions from the the valence to conduction surface
bands,
\begin{align}
\mathcal{H} &= \mathcal{H}_0 + \mathcal{H}_1\nonumber
\\
&=\sum\limits_{\mathbf{k}}\left(\sum\limits_{i=1}^4\epsilon_{\mathbf{k},i}a^{\dag}_{\mathbf{k},i}a_{\mathbf{k},i}+
\frac{e}{\hbar}A_2\mathbf{A}(t)\cdot\sum\limits_{i,j=1}^4
a^{\dag}_{\mathbf{k},i}\mathcal{D}_{ij}(\mathbf{k})a_{\mathbf{k},j}\right),
\end{align}
where $a_{\mathbf{k},i}$ are the band operators with energy
dispersion $\epsilon_{1,2}=\epsilon_{-}$ and
$\epsilon_{3,4}=\epsilon_{+}$, respectively. $\mathbf{A}(t)$ is the
vector potential. $\mathcal{D}(\mathbf{k})$ is the interband
transition operator,
\begin{equation}
\mathcal{D}(\mathbf{k}) = \left(
\begin{array}{cccc}
0 & 0 & \hat{e}_+ & 0\\
0 & 0 & 0 & \hat{e}_-\\
\hat{e}_- & 0  & 0 & 0\\
0 & \hat{e}_+ & 0 & 0
\end{array}
\right)\label{optical}
\end{equation}
where $\hat{e}_{\pm}=\hat{e}_x\pm i\hat{e}_y$. Here we have
neglected the small wavevector of light. Taking into account the
momentum conservation $\mathbf{k}$ for the initial $|i\rangle$ and
final $|f\rangle$ states, only the excitation processes pictured in
Fig.~1(a) contribute to the light absorption. The absorption power per
unit area is given by $W_a=\Gamma\hbar\omega$, where $\Gamma$ is the
absorption events per unit time per unit area, calculated by using
Fermi's golden rule to be
$\Gamma=(2\pi/\hbar)\sum_{i,f,\mathbf{k}}\left|\langle
f|\mathcal{H}_1|i\rangle\right|^2\delta(E_f-E_i-\hbar\omega)$, and
$\omega$ is the optical frequency. The incident energy flux $W_i$ is
given by $W_i=(c/4\pi)|\mathbf{A}(t)|^2\omega^2$. Then the
absorbance is defined as the ratio of absorbed light energy flux
divided by the incident light energy flux
\begin{equation}
\mathcal{P}_{\mathrm{gapless}} = \frac{W_a}{W_i} =
\frac{\pi}{2}\alpha,
\end{equation} where $\alpha$ is the fine structure
constant. The universal optical absorption does not depend on the
photon energy. For the interaction of light with relativistic
helical surface states is described by a single coupling constant,
i.e. the fine structure constant. The Fermi velocity is only a
prefactor for both $\mathcal{H}_0$ and $\mathcal{H}_1$, accordingly,
one can expect that the coefficient may not change the strength of
the interaction, as indeed our calculations show. When a TI film is
suspended in air, $\alpha=1/137$ and thus $\mathcal{P}=1.15\%$. It
is among materials with highest absorbance, about 200 times larger
than bulk $\mathrm{Hg_{1-x}Cd_xTe}$ of equivalent thickness, with absorption
coefficient about $100cm^{-1}$ for incident infrared
light\cite{dornhaus1976}. It is also just half of the value of a
single layer graphene. This is because for 3D TI film, the up and
down surfaces gives 2 gapless Dirac fermions, while for single layer
graphene, the two Dirac cones together and two real spins provide a
degenerate factor of 4.

As the energy gap is opened around the Dirac point in the thin film
TI like $\mathrm{Bi_2Se_3}$ less than 6 QLs, interband real
transitions between the conduction and valence surface bands can be
excited by optical field when the photon energy $\hbar\omega$
exceeds the surface band gap $2|\tilde{M}_0|$ (Fig.~1(b)) . Around the
band edge ($\mathbf{k}\rightarrow0$), the transition operator is exactly
the same as in the gapless case, and the absorbance is
\begin{equation}
\mathcal{P}_{\mathrm{gapped}} =
\pi\alpha\frac{1}{1+2\tilde{M}_0\tilde{B}_2/\tilde{A}_2^2}.
\end{equation}
Thus the absorbance for the gapped surface states is around
$\pi\alpha$, about twice higher than the gapless case. In
particular, for 2 and 3QL, the absorbance is $1.8\%$ and $2.05\%$,
smaller than $\pi\alpha$, and for 4 and 5QL, the absorbance is
$3.85\%$ and $6.22\%$, larger than $\pi\alpha$. The sign of
$\tilde{M}_0\tilde{B}_2$ determines the nontrivial or trivial
topology of the quasi-2D band structure\cite{bernevig2006d,Liu2010}.
$\tilde{M}_0\tilde{B}_2<0$ and
$\mathcal{P}_{\mathrm{gapped}}>\pi\alpha$, the system is in the
inverted regime, i.e. the QSH insulating states; while
$\tilde{M}_0\tilde{B}_2>0$ and
$\mathcal{P}_{\mathrm{gapped}}<\pi\alpha$, the system is just
conventional insulator. This suggests an optical way to identify the
topological nature other than transport
measurement\cite{koenig2007}.

In reality, the absorption occurs not just around band edge, so from
the point of view for application, we should get the full surface
absorption spectrum. Thus the transition rate is
\begin{equation}
\Gamma=(2\pi/\hbar)\sum_{i,f,\mathbf{k}}\left|\langle
f|\mathcal{H}_1|i\rangle\right|^2\delta(E_f-E_i-\hbar\omega)
\end{equation}
The absorbance is
\begin{widetext}
\begin{align}
\mathcal{P}_{\textrm{gapped}}(\omega) &=\frac{\Gamma_{\textrm{total}}(\omega)\hbar\omega}{W_i}
=\frac{\mathcal{P}_{\textrm{gapped}}(\tilde{M}_0)}{\sqrt{1+\frac{\tilde{B}_2^2(\hbar^2\omega^2-4\tilde{M}^2)}{\left(\tilde{A}_2^2+2\tilde{M}\tilde{B}_2\right)^2}}}
\left(1-\frac{\tilde{A}_2^2}{\tilde{B}_2^2}\frac{-(\tilde{A}_2^2+2\tilde{M}\tilde{B}_2)+
\sqrt{(\tilde{A}_2^2+2\tilde{M}\tilde{B}_2)^2+\tilde{B}_2^2(\hbar^2\omega^2-4\tilde{M}^2)}}{\hbar^2\omega^2}\right),
\end{align}
\end{widetext}
$\mathcal{P}_{\textrm{gapped}}(\tilde{M}_0)$ is the band edge
absorbance and the parameters for $\mathrm{Bi_2Se_3}$ are in Ref.~9. For 4 QLs $\mathrm{Bi_2Se_3}$, the
absorbance is shown in Fig.~2(a). We can see that the absorbance is
high in the allowed frequency range, indicating a good frequency
coverage when TI is used for photodetection. When the 3D TI is
reduced to quasi-2D, the absorbance is enhanced only near the band
edge, and when the photon energy increases, the absorbance drops to
the 3D value of $\mathcal{P}_{\mathrm{gapless}}=\pi\alpha/2$.

The gapped surface bands in the thin film TI can be classified into
two classes, with one class being the time reversal partner of the
other. Bands $1$ and $3$ are one class, the other is $2$  and $4$
(Fig.~2(b)). As shown in Eq.~(\ref{optical}), near the band edge,
right-handed ($\sigma_+$) light couples only to $1$ and $3$, while
left-handed ($\sigma_-$) light couples only to $2$ and $4$. It will
cause transition between the spin-down (up) valence state and the
spin-up (down) conduction state. For conventional semiconductors
like GaAs quantum well in [001] direction, the optical selection
rules from heavy hole bands to conduction bands are as
$\sigma_{\pm}: |S\rangle\otimes|\uparrow/\downarrow\rangle
\rightarrow\mp\frac{1}{\sqrt{2}}\left(|X\rangle\pm
i|Y\rangle\right)\otimes|\uparrow/\downarrow\rangle$. Here
$|S\rangle$, $|X\rangle$, $|Y\rangle$ and
$|\uparrow/\downarrow\rangle$ represents the orbital- and spin-part
wavefunction, respectively. We can see clearly in Fig.~2(b) that the
spin optical transition in the thin film TI is very different from
the conventional direct-gap semiconductor such as GaAs, where the
spin wavefunction remains unchanged. This unique
spin optical transition selection rules make the thin film TI attractive
for spintronics applications like fast switching.\cite{nishikawa1995}
\begin{figure}[htbp]
\begin{center}
\includegraphics[width=2.7in,clip=true]{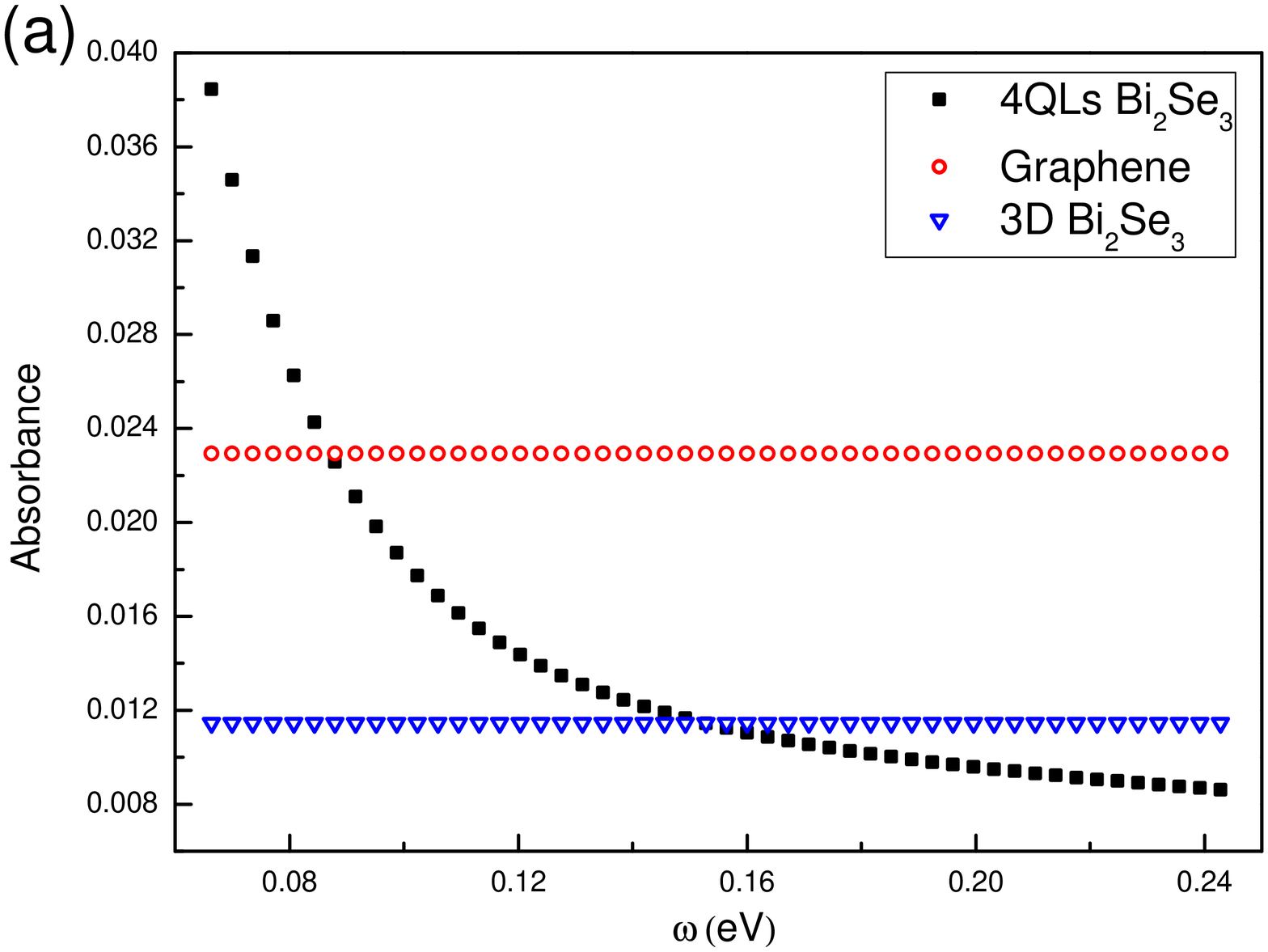}
\includegraphics[width=2.7in,clip=true]{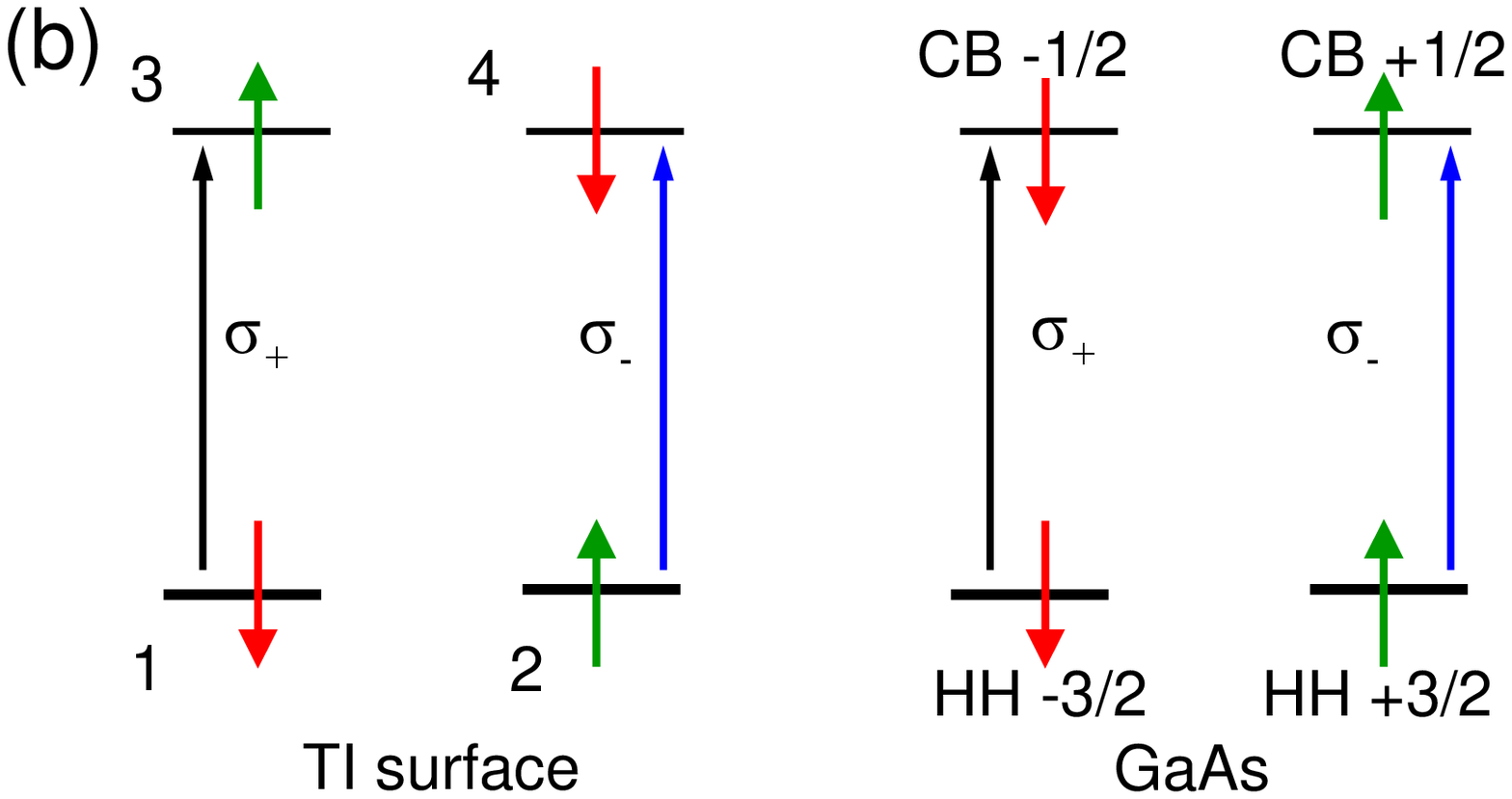}
\caption{\textbf{Absorption spectrum and selection rules.} (a)
Absorbance for graphene, 3D and 2D TI surface bands. (b) Comparison
of spin optical selection rules in GaAs and TI surface bands. The up
and down arrow denotes the spin up and spin down, respectively. CB
and HH denotes the conduction bands and heavy hole bands in GaAs.}
\end{center} \label{fig2}
\end{figure}

\section{TI-based photodetector}
Recently, graphene has been experimentally demonstrated as a high
performance THz $\sim$ Infrared photodetector due to its zero band
gap and high absorbance\cite{xiafn2009}. Similarly, the TI surface
may also be used for such application. The figure of merit of
photodetector is characterized by its SNR. Fig.~3(a) shows a typical
photodetector using the TI thin film - the photoconductive
resistor\cite{hamamatsu2004}. The particle and hole pairs will be
generated through light absorption and the system becomes more
electric conductive. Such change in conductivity can be easily
measured in an electronic circuit. The SNR in such system is defined
by $\mathrm{SNR}=I_{\mathrm{ph}}/\sigma_{I_{\mathrm{n}}}$, where
$I_{\mathrm{ph}}$ is the photocurrent and $\sigma_{I_{\mathrm{n}}}$
is standard deviation of the noise current whose major source is the
thermal noise (Johnson noise) at finite temperature $T$ when the
voltage bias $V \ll k_BT/q$ \cite{callegaro2006,noise2008}.
$I_{\mathrm{ph}}=2q\eta PS/E_{ph}$, here a factor of 2 represents
the electron-hole pair, $q$ is the electron charge, $\eta$ is the
quantum efficiency, $P$ is the incident light power density, $S$ is
the detecting area and $E_{ph}$ is the incident photon energy.
$\sigma_{I_n}=\sqrt{4k_BT\Delta f/R}$, whereas $\Delta f$ is the
bandwidth of the detector, inversely proportional to its integration
time and $R$ is the resistance. Thus the performance of a
photodetector rely on the intrinsic properties of the photodetecting
material, for $\eta$ is determined by the absorbance and $R$ is
determined by the resistivity $\rho$. The higher absorbance and the
larger resistivity (larger band gap) the material have, the better
it is to be used as a photodetector. A detailed calculation of the
SNR of TI photodetectors is provided in the supplementary
file\cite{epaps2010}. Due to the high absorbance, 3D TI can also be
used as a high performance THz $\sim$ Infrared ($0\sim0.3$eV)
photodetector, whose SNR is comparable than the commercially used
bulk $\mathrm{Hg_{1-x}Cd_{x}Te}$ (Fig.~3(b)). In particular, this
full spectrum can be covered by a single 3D TI detector, while for
$\mathrm{Hg_{1-x}Cd_{x}Te}$, multiple structures each covering a
segment of the bandwidth by tuning the fraction $x$ (and then band
gap) have to be used, to maintain high performance, which brings
much more complexities in fabrication. Moreover, the resistivity
$\rho$ can be greatly enhanced when surface opens a gap with the
thickness of thin film less than several QLs. Indeed, the SNR can be
improved to be about 15 times better than
$\mathrm{Hg_{1-x}Cd_{x}Te}$ (Fig.~3(b)). Besides this, the band gap
of such quasi-2D system are tunable form 5meV to 0.25eV by varying
its thickness\cite{zhangyi2010} (Fig.~3(b)) where the high energy
cut-off is 0.6eV here due to the confinement of bulk
bands\cite{zhangyi2010}. Photodetectors made from the
$\mathrm{Bi_2Se_3}$ family of TIs are potentially easier to
manufacture and can overcome the limitations of
$\mathrm{Hg_{1-x}Cd_{x}Te}$.
\begin{figure}[htbp]
\begin{center}
\includegraphics[width=3.3in,clip=true]{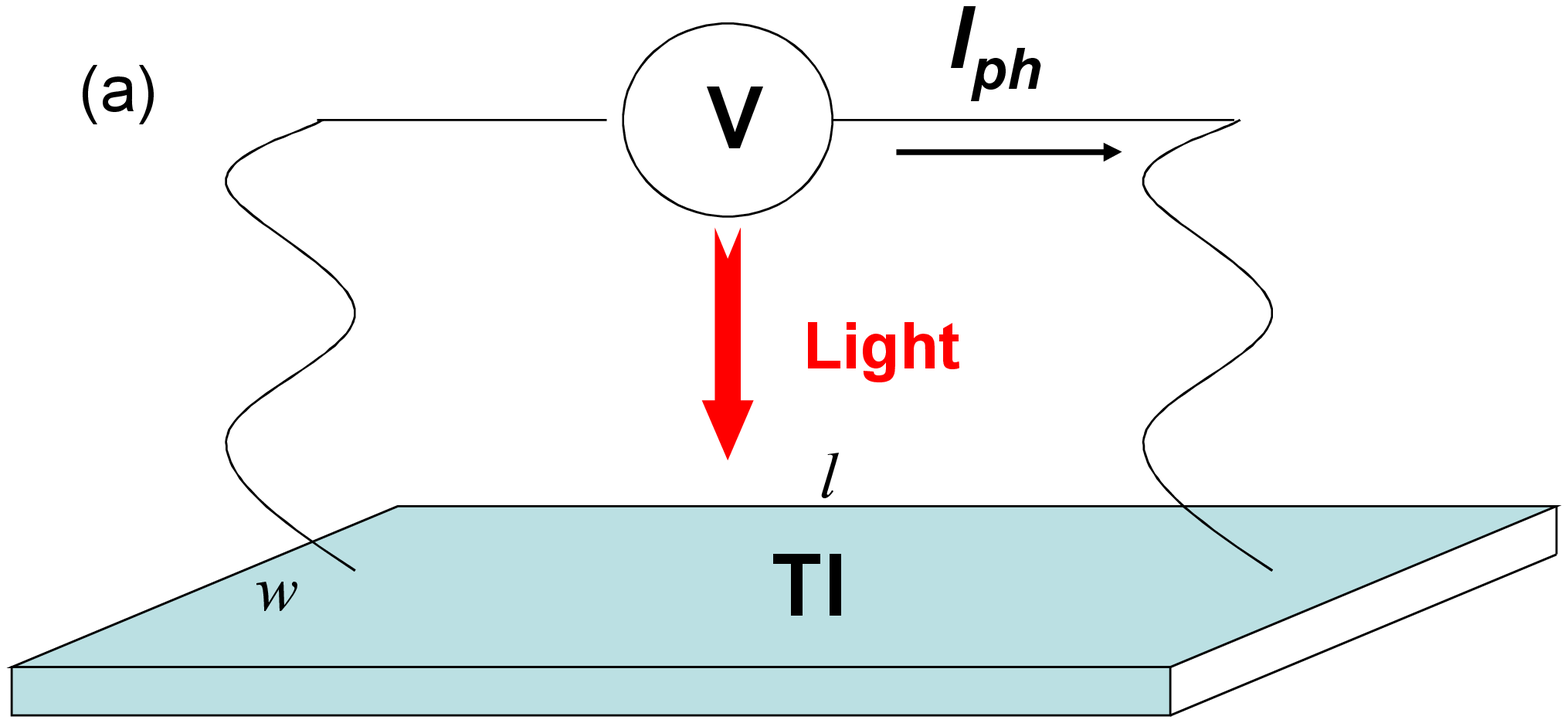}
\includegraphics[width=3.3in,clip=true]{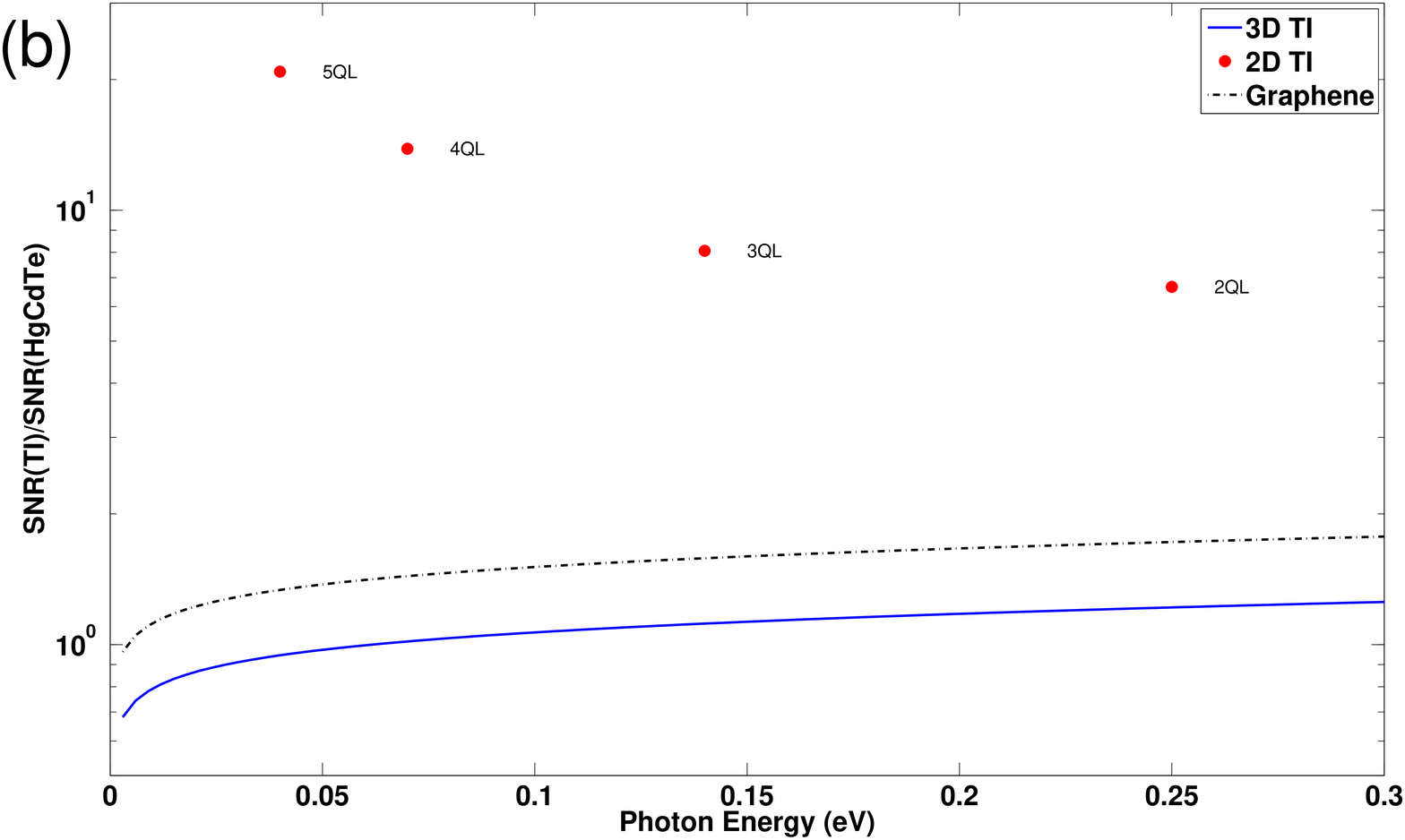}
\caption{\textbf{Photodetector device.} (a) Schematics of a
typical photo-resistor. (b) The SNR of 3D TI in full
spectrum (blue line), graphene (dashed line) and 2D TI at the band edge (red point) comparing with bulk
$\mathrm{Hg_{1-x}Cd_{x}Te}$.}
\end{center}\label{fig3}
\end{figure}

\section{Conclusion and outlook}
In summary, we discussed in this paper the fine structure constant
defined high optical absorption of TI and its unique selection
rules. Broadband and high performance photodetectors based on TI are
proposed. They may find a wide range of photonic applications
including thermal detection, high-speed optical communications,
interconnects, terahertz detection, imaging, remote sensing,
surveillance and spectroscopy\cite{xiafn2009}. In addition to
photodetection, TI may also be used in future for other
optoelectronic devices, such as terahertz laser\cite{dubinov2009},
waveguide\cite{mikhailov2007}, plasmon based radiation generation
and detection \cite{ryzhii2006} and transparent
electrode\cite{ksKim2009} etc. in analogy to graphene.

\begin{acknowledgments}
We wish to thank S. Raghu, J. Maciejko, S. B. Chung, R. D. Li and B.
F. Zhu for insightful discussions. J.W. acknowledge the support from
NSF of China (Grant No.10774086), and the Program of Basic Research
Development of China (Grant No. 2006CB921500).
\end{acknowledgments}

\end{document}